\documentclass[12pt]{article}
\usepackage{amsfonts}
\usepackage{amssymb}
\usepackage{graphics,psboxit,amsmath}
\usepackage{subfigure}
\usepackage{graphicx}
\usepackage{verbatim}
\usepackage{epic}

\def\hybrid{\topmargin 0pt \oddsidemargin 0pt 
        \headheight 0pt \headsep 0pt
        \textwidth 16,5cm 
        \textheight 23cm 
        \marginparwidth .875in
        \parskip 5pt plus 1pt \jot = 1.5ex}

\hybrid

\def\baselinestretch{1.2}

\catcode`\@=11

\def\marginnote#1{}
\newcount\hour
\newcount\minute
\newtoks\amorpm
\hour=\time\divide\hour by60
\minute=\time{\multiply\hour by60 \global\advance\minute by-\hour}
\edef\standardtime{{\ifnum\hour<12 \global\amorpm={am}%
        \else\global\amorpm={pm}\advance\hour by-12 \fi
        \ifnum\hour=0 \hour=12 \fi
        \number\hour:\ifnum\minute<10 0\fi\number\minute\the\amorpm}}
\edef\militarytime{\number\hour:\ifnum\minute<10 0\fi\number\minute}

\def\draftlabel#1{{\@bsphack\if@filesw {\let\thepage\relax
   \xdef\@gtempa{\write\@auxout{\string
      \newlabel{#1}{{\@currentlabel}{\thepage}}}}}\@gtempa
   \if@nobreak \ifvmode\nobreak\fi\fi\fi\@esphack}
        \gdef\@eqnlabel{#1}}
\def\@eqnlabel{}
\def\@vacuum{}
\def\draftmarginnote#1{\marginpar{\raggedright\scriptsize\tt#1}}

\def\draft{\oddsidemargin -.5truein
        \def\@oddfoot{\sl preliminary draft \hfil
        \rm\thepage\hfil\sl\today\quad\militarytime}
        \let\@evenfoot\@oddfoot \overfullrule 3pt
        \let\label=\draftlabel
        \let\marginnote=\draftmarginnote
   \def\@eqnnum{(\theequation)\rlap{\kern\marginparsep\tt\@eqnlabel}%
\global\let\@eqnlabel\@vacuum} }

\def\draft2{
        \def\@oddfoot{\sl preliminary draft \hfil
        \rm\thepage\hfil\sl\today\quad\militarytime}
        \let\@evenfoot\@oddfoot \overfullrule 3pt
        \let\label=\draftlabel
        \let\marginnote=\draftmarginnote
   \def\@eqnnum{(\theequation)\rlap{\kern\marginparsep\tt\@eqnlabel}%
\global\let\@eqnlabel\@vacuum} }

\def\preprint{\twocolumn\sloppy\flushbottom\parindent 2em
        \leftmargini 2em\leftmarginv .5em\leftmarginvi .5em
        \oddsidemargin -.5in \evensidemargin -.5in
        \columnsep .4in \footheight 0pt
        \textwidth 10.in \topmargin -.4in
        \headheight 12pt \topskip .4in
        \textheight 6.9in \footskip 0pt
        \def\@oddhead{\thepage\hfil\addtocounter{page}{1}\thepage}
        \let\@evenhead\@oddhead \def\@oddfoot{} \def\@evenfoot{} }

\def\numberbysection{\@addtoreset{equation}{section}
        \def\theequation{\thesection.\arabic{equation}}}

\def\underline#1{\relax\ifmmode\@@underline#1\else
        $\@@underline{\hbox{#1}}$\relax\fi}

\def\titlepage{\@restonecolfalse\if@twocolumn\@restonecoltrue\onecolumn
     \else \newpage \fi \thispagestyle{empty}\c@page\z@
        \def\thefootnote{\fnsymbol{footnote}} }

\def\endtitlepage{\if@restonecol\twocolumn \else \newpage \fi
        \def\thefootnote{\arabic{footnote}}
        \setcounter{footnote}{0}} 

\catcode`@=12
\relax

\def\figcap{\section*{Figure Captions\markboth
        {FIGURECAPTIONS}{FIGURECAPTIONS}}\list
        {Figure \arabic{enumi}:\hfill}{\settowidth\labelwidth{Figure
999:}
        \leftmargin\labelwidth
        \advance\leftmargin\labelsep\usecounter{enumi}}}
 \relax
\def\tablecap{\section*{Table Captions\markboth
        {TABLECAPTIONS}{TABLECAPTIONS}}\list
        {Table \arabic{enumi}:\hfill}{\settowidth\labelwidth{Table
999:}
        \leftmargin\labelwidth
        \advance\leftmargin\labelsep\usecounter{enumi}}}
 \relax
\def\reflist{\section*{References\markboth
        {REFLIST}{REFLIST}}\list
        {[\arabic{enumi}]\hfill}{\settowidth\labelwidth{[999]}
        \leftmargin\labelwidth
        \advance\leftmargin\labelsep\usecounter{enumi}}}
 \relax
\makeatletter
\newcounter{pubctr}
\def\publist{\@ifnextchar[{\@publist}{\@@publist}}
\def\@publist[#1]{\list
        {[\arabic{pubctr}]\hfill}{\settowidth\labelwidth{[999]}
        \leftmargin\labelwidth
        \advance\leftmargin\labelsep
        \@nmbrlisttrue\def\@listctr{pubctr}
        \setcounter{pubctr}{#1}\addtocounter{pubctr}{-1}}}
\def\@@publist{\list
        {[\arabic{pubctr}]\hfill}{\settowidth\labelwidth{[999]}
        \leftmargin\labelwidth
        \advance\leftmargin\labelsep
        \@nmbrlisttrue\def\@listctr{pubctr}}}
 \relax
\makeatother

\def\ba{\begin{equation}}
\def\ea{\end{equation}}

\def\d{\delta}
\def\D{\Delta}
\def\e{\epsilon}

\def\Th{\Theta}

\def\no{\noindent}

\def\qq{\qquad}

\def\IR{\relax{\rm I\kern-.18em R}}

\def \ha {{1\over 2}}

\def \ov {\over}

\def\diag{{\rm diag}}

\begin{document}


\renewcommand{\theequation}{\thesection.\arabic{equation}}
\csname @addtoreset\endcsname{equation}{section}

\newcommand{\eqn}[1]{(\ref{#1})}
\newcommand{\be}{\begin{eqnarray}}
\newcommand{\ee}{\end{eqnarray}}
\newcommand{\non}{\nonumber}
\begin{titlepage}
\strut\hfill
\vskip 1.3cm
\begin{center}


{\Large \bf Contracted and expanded integrable structures}

\vskip 0.5in

{\bf Anastasia Doikou}\phantom{x} and\phantom{x} {\bf Konstadinos Sfetsos}
\vskip 0.1in

Department of Engineering Sciences, University of Patras,\\
26110 Patras, Greece\\

\vskip .1in

\vskip .15in

{\footnotesize {\tt adoikou@upatras.gr},
\ \ {\tt sfetsos@upatras.gr}}\\

\end{center}

\vskip .4in

\centerline{\bf Abstract}

\no
We propose a generic framework to obtain certain
types of contracted and centrally extended algebras. This is
based on the existence of quadratic algebras (reflection
algebras and twisted Yangians), naturally arising in the context of boundary integrable models.
A quite old misconception regarding the ``expansion" of the
$E_2$ algebra into $sl_2$ is resolved using the
representation theory of the aforementioned quadratic algebras.
We also obtain centrally extended algebras
associated to rational and trigonometric ($q$-deformed) $R$-matrices
that are solutions of the Yang--Baxter equation.

\vfill
\no


\end{titlepage}
\vfill
\eject


\tableofcontents

\def\baselinestretch{1.2}
\baselineskip 20 pt
\no

\section{Introduction}

The original motivation for the present work comes from the following
interesting property of the non-semi-simple algebras that one obtains under Inon\"u--Wigner type
contractions. It looks as if there is a non-linear realization
of the original semi-simple algebra in terms of the generators of the non-semi-simple algebras, which
comes under the general name of expansion (see for instance \cite{gilmore}).
As a concrete example consider
the $E_2$ algebra corresponding to the three-dimensional Euclidean group.
This arises also as an Inon\"u--Wigner
contraction of the $sl_2$ and $su(2)$ algebras. The commutation rules are
\be
[J,\ P^\pm]=\pm P^\pm\ ,\qq [P^+,\ P^-]=0\ ,
\label{e22}
\ee
with the quadratic Casimir operator being $C=P^+P^-$.
Let us define
\be
Y^{\pm} = J P^{\pm}\ .
\label{exp0}
\ee
Then, it is straightforward to show that
\be
[J,\ Y^{\pm}] = \pm Y^{\pm} \ ,\qq [Y^+,\ Y^-] = -2 J P^+ P^- \ ,
 \label{exp1}
\ee
which are essentially the commutation relations of the $sl_2$ algebra.
Defining $\tilde Y^{\pm} ={Y^{\pm }\ov \sqrt{P^+P^-}}$,
then the elements $\tilde Y^{\pm},\ J$ seem to generate the $sl_2$ algebra. A similar
realization leads to the $su(2)$ commutation relations, that is $\tilde Y^{\pm} ={Y^{\pm }\ov \sqrt{-P^+P^-}}$.
The result is very appealing but it has a few drawbacks. From a physical point of view it is very difficult
to visualize how information defined in the two-dimensional infinite flat plane can be used to
reconstruct the curved manifold corresponding to the group spaces for $SU(2)$ and $SL(2)$.
In addition,
from a more mathematical view point we can readily check that it is only the identity of the corresponding
Lie-algebras that can be expressed in terms of $E_2$ representations.

\no
We will show that this so called expansion is only an apparent one.
In particular, the basic difference is already encoded from the algebraic point of view
in the co-product. For the generators of $sl_2$ this is given by
\be
\Delta (X) = {\mathbb I} \otimes X + X \otimes {\mathbb I}\ , \qq
X \in \{J,\ J^{\pm} \}\ . \label{co1}
\ee
Then one naively computes that
\be
\D(Y^\pm) = Y^\pm \otimes {\mathbb I} + {\mathbb I} \otimes Y^\pm + J \otimes P^\pm +
P^\pm \otimes J \ ,
\label{co2}
\ee
which already suggests that, contrary to what one would have expected $\Delta(Y^{\pm})$ and $\D(J)$
do not belong to $U(sl_2) \otimes U(sl_2)$
--notice in (\ref{co2}) one borrows elements from $E_2$ in order to construct the co-product--,
although they still satisfy \eqn{exp1},
so one should better search for a broader algebra. Indeed we show here that the
associated extended algebra in this case is
the contracted $sl_2$ twisted Yangian (see e.g. \cite{sklyanin, molev, doikou0, moras}),
which rules integrable models
with non-trivial integrable boundary conditions. In other words we use the notion
of integrability in order to show that the
``expansion'' presented in \cite{gilmore} is not valid.

\no
Moreover, we show that the symmetry breaking mechanism due to the presence
of appropriate boundary conditions may be also
exploited in order to obtain centrally extended algebras via suitable contraction procedures.
We also use the boundary algebra to
obtain the relevant Casimir. This is perhaps the simplest and most straightforward means
to obtain the Casimir of usual and
deformed Lie algebras. One of the main points of this investigation is that
we are able to show that the associated open transfer matrix commutes
with the elements of the emerging contracted algebra.
We study here the simplest case, that is the $E_{2}^c$ algebra,
in order to illustrate the procedure followed, however this description
may be generalized for more complicated algebraic structures. Such an exhaustive analysis
however is beyond the intended scope
of the present work.

\no
Let us now outline the content of the present article. In section 2 we introduce
the fundamental quadratic algebras
that rule integrable models with periodic and non-trivial boundary conditions, that is
we review relevant aspects of the Yang-Baxter equation
\cite{baxter} and the associated quadratic relations (see e.g. \cite{tak, fad}) hich give sire to the quantum
algebras \cite{jimbo}--\cite{cha}.
In a similar spirit we also review the reflection algebra and
the twisted Yangian \cite{cherednik, sklyanin, moras}. In section 3 we examine
the contracted $sl_2$ twisted Yangian. We show, using representation theory of the relevant algebra,
that the ``expansion'' presented in the literature \cite{gilmore} is not the $sl_2$,
but the associated $E_2$ twisted Yangian. Next we exploit symmetry breaking mechanisms
due to the presence of suitable integrable
boundary conditions in order to obtain the centrally extended $E_{2}^c$ algebra.
In section 5 we construct the $q$ deformed version
of the $E_{2}^c$ algebra with the help of $U_q(sl_2) \otimes u(1)$ boundary symmetry.
In the last section a discussion on possible
directions for further study is presented.

\section{Quadratic algebras}

In this section we give a short review of the fundamental quadratic algebraic relations,
ruling the quantum integrable models, that is the
Yang--Baxter and reflection equations.

\no The Yang--Baxter equation \cite{baxter} is defined as
\be
R_{12}(\lambda_{1} -\lambda_{2})\ R_{13}(\lambda_{1})\
R_{23}(\lambda_{2})\ =\ R_{23}(\lambda_{2})\ R_{13}(\lambda_{1})\
R_{12}(\lambda_{1}-\lambda_{2}). \label{ybe2}
\ee
acting on
${\mathbb V}^{\otimes 3}$, and $R \in \mbox{End}({\mathbb
V}^{\otimes 2})$ $~R_{12} = R \otimes {\mathbb I}$, $R_{23} =
{\mathbb I} \otimes R$. From a physical view point it is well known
that the Yang--Baxter equation describes the factorization of
multi-particle scattering in integrable models (see e.g.
\cite{yang2}-\cite{korepin}). Given an $R$ matrix, as a solution of
the Yang--Baxter equation, we introduce the following fundamental
algebraic relations \cite{tak}, which essentially define a
particular algebra ${\cal A}$ (see e.g. \cite{tak})
\be
R_{12}(\lambda_1-\lambda_2)\ L_1(\lambda_1)\ L_2(\lambda_2)\ =\
L_2(\lambda_2)\ L_1(\lambda_1)\ R_{12}(\lambda_1-\lambda_2)\ ,
\label{fundam}
\ee
where $L \in \mbox{End}({\mathbb V}) \otimes
{\cal A}$, with ${\cal A}$ being the algebra defined by
\eqn{fundam}. This allows the construction of tensorial
representations of the later algebra as
\be
T_a(\lambda) \ =\
L_{aN}(\lambda-\theta_N)\ L_{a N-1}(\lambda-\theta_{N-1}) \ldots
L_{a2}(\lambda-\theta_2)\ L_{a1}(\lambda-\theta_1)\ , \label{mono}
\ee
where $T(\lambda) \in \mbox{End}({\mathbb V}) \otimes {\cal
A}^{\otimes N}$.
For historical reasons, the $a$ space is called
``auxiliary", whereas the spaces $1, \dots, N$ are called
``quantum''. For simplicity we usually suppress all quantum spaces
when writing down the monodromy matrix. Also $\theta_i$ are free
complex parameters and are called inhomogeneities. Using the
fundamental algebra (\ref{fundam}) one may show that
\be
\Big [ tr
T(\lambda),\ tr T(\mu) \Big ] =0\ ,
\ee
where $tr T(\lambda) \in
{\cal A}^{\otimes N}$ and the trace is taken over the auxiliary
space. The latter relation guarantees the integrability of the
system. Once the algebra ${\cal A}$ is represented, the tensorial
representation acquires the meaning of the monodromy matrix of a
quantum spin chain and and $tr T$, the corresponding transfer
matrix, may be diagonalized using for instance Bethe ansatz
techniques.

\no
We shall now introduce the reflection equation \cite{sklyanin, cherednik}.
In fact we shall consider two types of equations associated to two distinct algebras, the reflection
algebra \cite{sklyanin} and the twisted Yangian \cite{molev, moras}. These two algebras describe essentially
the algebraic content of integrable models with two distinct types of boundary conditions known as soliton
preserving (SP) and soliton non-preserving (SNP), respectively. A spin chain-like system with SNP
boundary conditions was first derived in \cite{doikou0},
whereas generalizations were studied in \cite{annecy1, annecy2}.

\no
{\bf Reflection algebra}: The equation associated to the reflection algebra ${\cal R}$
(SP boundary conditions) is given by \cite{cherednik, sklyanin}
\be
R_{12}(\lambda_{1} -\lambda_{2})\ {\mathbb K}_{1}(\lambda_{1})\ R_{21}(\lambda_{1}
+\lambda_{2})\ {\mathbb K}_{2}(\lambda_{2})=
{\mathbb K}_{2}(\lambda_{2})\ R_{12}(\lambda_{1} +\lambda_{2})\
 {\mathbb K}_{1}(\lambda_{1})\ R_{21}(\lambda_{1} -\lambda_{2})\ ,
\label{re2}
\ee
acting on ${\mathbb V}^{\otimes 2} $ and as customary we follow the notation
${\mathbb K}_{1} ={\mathbb K} \otimes {\mathbb I}$ and $ {\mathbb K}_{2} =
{\mathbb I} \otimes {\mathbb K}$.
Also $R_{21} = {\cal P}\ R_{12}\ {\cal P}$, where ${\cal P}$ is the permutation operator: ${\cal P}
(a \otimes b)= b \otimes a$ and also ${\mathbb K} \in \mbox{End}({\mathbb V})\otimes {\cal R}$.

\no
{\bf Twisted Yangian}: The twisted Yangian (SNP boundary conditions) ${\cal T}$ defined
by \cite{molev, moras}
\be
R_{12}(\lambda_{1} -\lambda_{2})\ \tilde {\mathbb K}_1(\lambda_1)\ \hat R_{12}(\lambda_{1}
 +\lambda_{2})\ \tilde
{\mathbb K}_{2}(\lambda_{2})= \tilde {\mathbb K}_{2}(\lambda_{2})\ \hat R_{12}(\lambda_{1}
 +\lambda_{2})\ \tilde
{\mathbb K}_{1}(\lambda_{1})\ R_{12}(\lambda_{1} -\lambda_{2}) \ ,
\label{ty}
\ee
where
\be
\hat R_{12}(\lambda) = R_{12}^{t_1}(-\lambda -i\rho)\ ,
\label{ty0}
\ee
with $\tilde {\mathbb K} \in \mbox{End}({\mathbb V}) \otimes {\cal T}$ and $^{t_1}\!$
denotes transposition on the first space.

\no
In general the representations of the later algebras may be expressed, in an index free notation,
as \cite{sklyanin, molev, doikou0}
\be
\mbox{Reflection algebra}:
&& {\mathbb K}(\lambda) = L(\lambda-\Theta)\ K(\lambda)\ L^{-1}(-\lambda- \Theta)\ ,
\non\\
\mbox{Twisted Yangian} : && \tilde {\mathbb K}(\lambda) = L(\lambda-\Theta)\
\tilde K(\lambda)\ L^{t}(-\lambda-\rho\ - \Theta)\ , \label{kk1} \ee
where the matrices $K,\ \tilde K$ are $c$-number representations of
the aforementioned algebras, $\Theta$ is an inhomogeneity, and
$\rho$ is a constant that depends on the underlying algebra. For
instance, in the $gl_n$ case $\rho = {n \over 2}$. Tensor
representations of these algebra are given by
\be \mbox{Reflection
algebra}:
&& {\mathbb T}_0(\lambda)= T_0(\lambda)\ K_0(\lambda)\
T_0^{-1}(-\lambda)\ ,
\non\\
\mbox{Twisted Yangian}: && \tilde {\mathbb T}_0(\lambda)=
T_0(\lambda)\ \tilde K_0(\lambda)\ T_0^{t_0}(-\lambda -i\rho)\ ,
\label{tt1} \ee where we recall that $T$ is defined in (\ref{mono}).

\no
Let us now define the $N$ ``particle'' transfer matrix
\be
t(\lambda) = tr\{K^+(\lambda) {\mathbb T}(\lambda)\}\ , \qq
\tilde t(\lambda) = tr\{ \tilde K^+(\lambda) \tilde {\mathbb T}(\lambda)\}\ .
\label{transfer}
\ee
Clearly, for the one ``particle'' construction
we have ${\mathbb T} \to {\mathbb K}$ and $ \tilde {\mathbb T} \to \tilde {\mathbb K}$.
Also the $K^+, \tilde K^+$ matrices are $c$-number solutions of the reflection algebras
and twisted Yangian, respectively.
With the help of the quadratic exchange relations one
may show that (see e.g. \cite{sklyanin, doikou0})
\be
[t(\lambda),\ t(\mu) ] =0\ , \qq [\tilde t(\lambda),\ \tilde t(\mu) ] =0\ .
\label{coons}
\ee
In the following we shall mainly consider the $gl_n$ Yangian ${\cal Y} (gl_n)$ case.
The associated $R$- and $L$-matrices are then given by \cite{yang}
\be
R(\lambda) = {\mathbb I} + {i\over \lambda} {\cal P}\ ,
\qq L(\lambda)= {\mathbb I} + {i\over \lambda} {\mathbb P}\ ,
\label{def1}
\ee
where the entries ${\mathbb P}_{ab} \in gl_n$.

\section{$sl_2$ quadratic algebras; contractions and expansions}

In this section we shall focus on the situation associated to the ${\cal Y}(sl_2)$
$R$-matrix \cite{yang}. Then
the $L$-operator in (\ref{def1}) is explicitly given by
\be
L(\lambda)= {\mathbb I} + {i\over \lambda} \left(
\begin{array}{cc}
J+{1\over 2} &-J^- \\
J^+ &-J +{1\over 2}
 \\ \end{array} \right)\ ,
\label{ppp}
\ee
with $J,\ J^{\pm}$ the generators of $sl_2$ satisfying
\be
[J,\ J^{\pm}] = \pm J^{\pm}\ , \qq [J^+, J^-] = -2 J\ .
\label{sl}
\ee
Recall that the Casimir operator for $sl_2$ is $C = -J^2 + {1 \over 2}(J^+J^- + J^- J^+)$.
\footnote{Note that there exist a representation $\pi$
and a homomorphism $h$ defined as $\pi: su_2 \hookrightarrow \mbox{End}({\mathbb C}^2)$
and $h: su_2 \hookrightarrow sl_2$ such that
\be
\pi(J) = {\sigma^z \over 2}\ ,
\qq \pi(J^+) = \sigma^+\ , \qq \pi(J^-) = \sigma^- \non\\ h(J) = J\ , \qq
h(J^+) = J^+\ , \qq h(J^-) =-J^-\ .
\ee
Taking into account the above we conclude that
 \be
&& (\pi \otimes h) \Delta({\mathrm x})\ L(\lambda) = L(\lambda)\ (\pi \otimes h) \Delta({\mathrm x})
 \non\\
&&\Delta({\mathrm x})
 = {\mathbb I} \otimes {\mathrm x} + {\mathrm x} \otimes {\mathbb I} \ ,\qq {\mathrm x}\in su_2.
\ee }
As is well known, one may easily obtain from $sl_2$ by an Inon\"u--Wigner contraction the $E_2$ algebra.
Indeed setting $J^{\pm} = {1\over \epsilon} P^{\pm}$ with $\epsilon \to 0$
we have from (\ref{sl}) that
\be
[J,\ P^{\pm}] = \pm P^{\pm}\ , \qq [P^+, P^-] = 0\ .
\ee
The associated Casimir operator is $C= P_+\ P_-$.

\no
Let us point out that in the $sl_2$ case the reflection algebra coincides essentially
with the twisted Yangian due to the fact that $sl_2$ is self-conjugate. Equivalently, the
 $sl_2$ $L$-operator and the $R$-matrix are crossing symmetric, i.e.
\be
\sigma^y\ L^t(-\lambda -i)\ \sigma^y = L(\lambda)\ ,
\qq \sigma_1^y\ R_{12}^{t_1}(-\lambda -i)\ \sigma_1^y = R_{12}(\lambda)\ .
\ee
Moreover, one can easily check that, due to the fact that ${\mathbb P}^2 $ is proportional to the
Casimir operator, we have
\be
L^{-1}(-\lambda) \propto L(\lambda)\ .
\ee
Taking into account the representations of the reflection algebra and the twisted Yangian and the above relations,
it is straightforward to see that the two algebras coincide.

\no
In general the twisted Yangian can be defined up to a gauge transformation, that is $\hat L \to \ V \hat L V$, where $V^2 =1$.
Then the defining relations of the twisted Yangian are slightly modified after $\hat R$ is also redefined.
Indeed, by multiplying equation (\ref{ty}) from the right with $V_1\ V_2$, we end up with
\be
R_{12}(\lambda_{1} -\lambda_{2})\ \bar {\mathbb K}_1(\lambda_1)\
\bar R'_{12}(\lambda_{1} +\lambda_{2})\ \bar {\mathbb K}_{2}
(\lambda_{2})= \bar {\mathbb K}_{2}(\lambda_{2})\ \bar R_{12}(\lambda_{1}
+ \lambda_{2})\ \bar {\mathbb K}_{1}(\lambda_{1})\
R'_{12}(\lambda_{1} -\lambda_{2}) \ ,
\label{ty2}
\ee
where we define
\be
\bar {\mathbb K}= \tilde {\mathbb K}\ V\ , \qq
\bar R_{12}(\lambda)=V_2R^{t_1}_{12}(-\lambda-i\rho)V_2\ ,
\qq A'_{12} =V_1 V_2 A_{12} V_1 V_2\ .
\ee
A representation of the algebra defined in (\ref{ty2}) is
\be
\bar {\mathbb K} = L(\lambda -\Theta)\ \hat L(\lambda+\Theta)\ ,
\qq \mbox{where} \qq \hat L(\lambda) = V\ L^t(-\lambda -i\rho)\ V\ .
\ee
For simplicity we set henceforth $\Theta={i \over 2}$.

\no
Although we have seen that in this case the reflection algebra and twisted
Yangian coincide,
we will preserve the terminology to basically distinguish
two different types of boundary conditions, which are discussed below.

\subsection{The twisted Yangian and its contraction}

Here we shall consider\footnote{In many of the algebraic
manipulations, such as in the one leading to the second equality
below, we omit constant overall factors since from the relevant
defining equations either they drop out or they produce an
unimportant for our considerations overall factor.}
\be
\hat L(\lambda) = \sigma^x\ L^t(-\lambda -i)\ \sigma^x =1 + {i\ov \lambda}
\left(
  \begin{array}{cc}
    J+\ha & J^- \\
    -J^+ & -J + \ha \\
  \end{array}
\right) \ , \label{e31}
\ee
where we have used \eqn{def1} with
\eqn{ppp}. We form the generating function
\be
\bar {\mathbb K}
(\lambda) = L(\lambda-{i\over 2})\ \hat L(\lambda+{i\over 2})\ ,
\ee
where from \eqn{ppp} and \eqn{e31}
\be
 && L(\lambda-{i\over 2}) =
{\mathbb I} + {i\over \lambda}\ {\mathbb P} \ ,\qq {\mathbb P} =
\left( \begin{array}{cc}
J &-J^- \\
J^+ &-J \\ \end{array} \right )\ ,
\nonumber\\
&& \hat L(\lambda+{i\over 2}) = {\mathbb I} + {i\over \lambda}\ \hat
{\mathbb P} \ ,\qq \hat {\mathbb P} = \left( \begin{array}{cc}
J +1 &J^- \\
-J^+ &-J +1 \\ \end{array} \right )\ .
\ee
To obtain the charges in
involution we expand it in powers of ${1\over \lambda}$ as
\be
\bar
{\mathbb K} (\lambda) = {\mathbb I} + {1\over \lambda}\bar {\mathbb
K}^{(0)} + {1\over \lambda^2} \bar {\mathbb K}^{(1)}\ , \ee where
\be \bar {\mathbb K}^{(0)} = i \left(
\begin{array}{cc}
2J +1 &0 \\
0 &-2J+1 \\ \end{array} \right )\ ,
 \quad
\bar {\mathbb K}^{(1)} =\left(
\begin{array}{cc}
-J^2 -{1\over 2} \{J^+,\ J^-\} -2J &- 2J J^- \\
- 2 J J^+ & - J^2 -{1\over 2} \{J^+,\ J^-\}+ 2J
      \\ \end{array}
\right )\ . \non
\ee
Taking the trace we end up with
\be
\bar
t(\lambda) = tr\{\bar {\mathbb K}(\lambda) \}= {\mathbb I} + {i\over
\lambda}+ { \bar t^{(1)} \over \lambda^2}\ , \qq \mbox{where} ~~~~~
\bar t^{(1)} \propto J^2 +{1 \over 2}\{J^+,\ J^- \}\ . \label{cons1}
\ee
Here
$\bar t^{(1)}$ is the only non-trivial conserved quantity as is dictated by the
commutation relations \eqn{coons} because the expansion stops at $1\over \lambda^2$.
If we had to deal with higher terms in the expansion we would have more
conserved quantities (higher Casimir operators) as will be transparent in the subsequent sections,
when dealing with higher rank algebras.
This quantity will become the quadratic Casimir
of $E_2$ after contraction.

Notice that the conserved
quantity is not the $sl_2$ Casimir operator (it is structurally an
$su_2$-like Casimir), but rather an element of the abelian part of
the twisted Yangian. The reason is that the symmetry of the
particular boundary model is not an $sl_2$ one, but simply $u(1)$ as
dictated by the form of $\bar {\mathbb K}^{(0)}$. This is in
accordance with \cite{doikouy, doikoucrampe},
\be
[tr\{\bar {\mathbb
K}(\lambda)\},\ \bar {\mathbb K}_{ab}^{(0)} ] =0\ ,
\ee
a relation
that defines the exact symmetry of the transfer matrix. The transfer
matrix that enjoys the full $sl_2$ symmetry will be presented
subsequently.

\no
Let us define
$\bar {\mathbb K}_{12}^{(1)} = -2Y^-$ and $\bar {\mathbb K}_{21}^{(1)} = -2Y^+ $.
After performing the Inon\"u--Wigner contraction $J^{\pm} \to {1 \over \epsilon} P^{\pm}$, with $\e\to 0$,
we end up with
\be
Y^{\pm} = J P^{\pm}\ ,
\ee
precisely as in \eqn{exp0}. Therefore we have the exchange relations \eqn{e22} and seemingly
it looks as if one can expand $E_2$ back to $sl_2$ (see \cite{gilmore}). However, this is not true,
since $J$ and $Y^{\pm}$ are elements of an extended algebra,
i.e. the contracted $sl_2$ twisted Yangian ${\cal T}$
with exchange relations dictated by the quadratic equation (\ref{ty2}).
This will be more transparent in the following when constructing the $N$-tensor representation
of the twisted Yangian.

\subsubsection{ The $N$-particle construction}

In this section we basically illustrate the presence of non trivial co-products
further manifesting the existence of the underlying  non trivial algebra that is the twisted Yangian of $E_2$.
We define the $N$-tensor representation of the twisted Yangian as
\be
\bar {\mathbb T}_0(\lambda) = L_{0N}(\lambda-{i \over 2})
 \ldots L_{01}(\lambda-{i \over 2})\ \hat L_{01}(\lambda+{i \over 2}) \ldots
\hat L_{0N}(\lambda+{i \over 2})\ .
\ee
As before, an expansion in
powers of ${1 \over \lambda}$ leads to
\be \bar {\mathbb T}(\lambda)
= {\mathbb I} + \sum_{k=1}^{2N}{\bar {\mathbb T}^{(k-1)} \over
\lambda^{k}} \ , \label{exppa} \ee with \be \bar {\mathbb
T}_0^{(0)} & = & \sum_{i=1}^N \Big ( {\mathbb P}_{0i}+ \hat {\mathbb
P}_{0i} \Big )\ ,
\nonumber\\
\bar {\mathbb T}_0^{(1)} & = & -\Big (\sum_{i>j} {\mathbb P}_{0i}\ {\mathbb P}_{0j}
+ \sum_{i<j} \hat {\mathbb P}_{0i}\ \hat
{\mathbb P}_{0j}
 + \sum_{i, j} {\mathbb P}_{0i}\ \hat {\mathbb P}_{0j}\Big )\ \label{first0}
\ee
and so on.
The first non-trivial conserved quantity in this case is obtained after taking the trace of $\bar {\mathbb T}^{(1)}$.
One finds that
\be
\bar t^{(1)} \propto \sum_{i=1}^N \Big (J_i^2 +{1\over 2}
\{J_i^+,\ J_i^- \} \Big ) +4 \sum_{i < j} J_i\ J_j\ .
\ee
After performing the contraction the first conserved quantity is given by
\be
\bar t^{(1)} = \sum_{i=1}^N P^{+}_i P^-_i\ .
\ee
Notice that for $N=1$ one simply obtains the expected $E_2$ Casimir emerging directly from (\ref{cons1})
after contracting and keeping the highest order contribution. All quantities $\bar t^{(k)}$ (in the co-product form now)
form, as
dictated by the integrability condition (\ref{coons}), an abelian algebra (family of commuting operators), which is
part of the twisted Yangian of $E_2$.
It is worth noting that after contraction we consistently keep only
the highest order terms for each $\bar t^{(k)}$. This will be explained in more detail in section 4.

For the non-diagonal elements we have
$\bar {\mathbb T}_{12}= - 2{\mathbb Y}^-$
and $\bar {\mathbb T}_{21}= - 2{\mathbb Y}^+$, where
\be
{\mathbb Y}^\pm =\sum_{i=1}^N J_i P_i^\pm + 2 \sum_{i<j} J_i P_j^\pm \ ,
\label{cc}
\ee
with
the subscript standing for the $i^{th}$ site in the $N$ co-product sequence.
Recall that the co-product for all generators of $sl_2$ is given by (\ref{co1}).
On the other hand from (\ref{cc}) it is obvious that the two co-product of the underlying algebra,
given by the two particle case, is
\be
\Delta(Y^{\pm}) = Y^{\pm} \otimes {\mathbb I} + {\mathbb I} \otimes Y^{\pm} + 2 J \otimes P^{\pm} \ .
\label{c2}
\ee
Comparison with (\ref{co1}) suggests that $Y^{(\pm)},\ J$ belong to a broader deformed algebra,
which in the particular case is the contracted $sl_2$ twisted Yangian.
Notice the extra term appearing in the co-product,
which suggests that we deal with a deformed algebra, that is the contracted $sl_2$ twisted Yangian,
and not an extension to $sl_2$ as was claimed for instance in \cite{gilmore}.
Study of representations shows inconsistencies, i.e. $\Delta(Y^{\pm}),\ \Delta(J)$
do not belong to $U(sl_2) \otimes U(sl_2)$ as one might have expected,
but to a broader algebra the ${\cal T} \otimes {\cal A}$ (${\cal T}$ is a co-ideal of ${\cal A}$,
see also e.g.
\cite{dema, doikouy, doikou, myhecke}).
Furthermore, these co-products do not satisfy (\ref{e22}),
contrary to the naive co-products (\ref{co2}). If they were this would have been surprising
due to the fact that for one ``particle''
the expansion stops at order ${1 \over \lambda^2}$ so the relevant exchange
relations emanating from (\ref{ty2}) are somehow truncated.
However, once we consider the $N$ ``particle'' representation the
expansion involves higher orders, and thus the
associated exchange relations become more involved.

\subsection{The reflection algebra and its contraction}

We consider the representation ${\mathbb K}(\lambda)$ (\ref{kk1}) of
the reflection algebra with $\Theta = {i\over 2}$. The transfer
matrix is given by (\ref{transfer}). By choosing $K =
\mbox{diag}(-1,\ 1)$ and $K^+ = \mbox{diag}(1,\ -1)$ we are dealing
with a situation similar to the description above. We obtain the
same conserved charges in involution, and also the associated
transfer matrix enjoys the $u(1)$ symmetry. This is due the
aforementioned equivalence of the twisted-Yangian and reflection
algebras for $sl_2$.

\no
Consider now both $K = K^+ ={\mathbb I}$ in the transfer matrix $t$ (\ref{transfer}),
that is we
choose different boundary conditions for the associated physical system.
We consider directly the $N$-particle Hamiltonian ($N$-tensor representation),
 (\ref{tt1}) and (\ref{mono})
with $\theta_i =-{i\over 2}$, that is we have
\be
{\mathbb T}_0(\lambda) = L_{0N}(\lambda+{i \over 2})
 \ldots L_{01}(\lambda+{i \over 2})\ L_{01}(\lambda-{i \over 2}) \ldots
L_{0N}(\lambda-{i \over 2})\ . \ee As before, an expansion in powers
of ${1 \over \lambda}$ leads to \be t^{(1)} \propto \sum_{i=1}^N
\Big (J_i^2 -{1\over 2} \{J_i^+,\ J_i^- \} \Big ) +2 \sum_{i < j}
\left(2 J_i J_j - J_i^+ J^-_j - J_i^- J^+_j\right)\ , \ee
after we take the trace of the $1/\lambda^2$ term. Note the natural
appearance of the quadratic Casimir operator for $sl_2$ for the
1-particle case.

\no
In this case after contraction we obtain
\be
t^{(1)} \propto \sum_{i=1}^N P^+_i P^-_i + 2 \sum_{i< j} P_{i}^+ P_j^-\ .
\ee
Also, we can safely say that
\be
I = \sum_{i < j} P_i^- P^+_j \ ,
\ee
is also a conserved quantity because $P_i^- P_i^+$
is the Casimir associated at each site and commutes with all elements
acting on the same site.
It is worth noting that although both conserved quantities coincide in the one ``particle'' situation,
they are obviously different
when more ``particles''
are involved, $\bar t^{(1)} \neq t^{(1)}$, that is their co-products are different.
In this case the transfer matrix enjoys the full $sl_2$ symmetry
and it is natural that we obtain the $sl_2$
Casimir as a conserved quantity from the transfer matrix expansion.

\section{The $E_{2}^c$ extended algebra }

In the present section we aim at constructing the centrally extended $E_{2}^c$ algebra.
To achieve this we start from the $gl_3$
spin chain and break the symmetry down to $sl_2 \otimes u(1)$
by implementing appropriate boundary conditions.
It has been known \cite{done} that by implementing appropriate boundary conditions
one can break the $gl_n$ symmetry of a spin chain model to $gl_l \otimes gl_{n-l}$,
where $l$ is an integer depending on the
choice of boundary. We shall exploit this phenomenon in order
to perform a contraction of the boundary algebra to $E_{2}^c$.

\no
The $gl_n$ algebra is
\be
[J_{ij},\ J_{kl}] = \d_{il} J_{kj}-\d_{jk} J_{il}\ ,\qq i=1,2,\dots , n\ .
\ee
It is generated by
\be
 J^{+(i)}= J^{i+1\ i}\ ,\qq J^{-(i)}= J^{i\ i+1}\ ,\qq e^{(i)} = J^{ii}\ .
\ee
Define $\displaystyle s^{(k)} = e^{(k)} -e^{(k+1)}$, then the following commutation relations are valid
\be
[J^{+(k)},\ J^{-(l)}] = \delta_{kl} s^{(k)}\ , \qq
[s^{(k)},\ J^{\pm (l)}] =\pm (2 \delta_{kl} - \delta_{k\ l+1} - \delta_{k\ l-1}) J^{\pm (l)}\
\label{sllk}
\ee
and $\displaystyle \sum_{i=1}^n{e^{(i)}}$ belongs to the center of the algebra.

\no
From now on we focus on the case of $gl_3$. The $L$ matrix is expressed as in (\ref{def1}),
where ${\mathbb P}$ in term of the $gl_3$ elements takes the form
\be
{\mathbb P} = \left(
\begin{array}{ccc}
e^{(1)} &J^{-(1)} &\Lambda^{+}\\
J^{+(1)} &e^{(2)} &J^{-(2)} \\
\Lambda^{-} &J^{+(2)} &e^{(3)} \\ \end{array} \right )\ ,\quad
\mbox{where}\qq \Lambda^{\pm} = \pm[J^{\pm(1)},\ J^{\pm(2)} ]\ .
\label{def3}
\ee
Consider next the $N$-tensor representation of the
reflection algebra (\ref{tt1}), (\ref{mono}), with $\theta_i=0$ and
$L$ given in (\ref{def1}) and (\ref{def3}). We choose as $K$ the
following diagonal matrix (for a more general solution see
\cite{deve, done})
\be
K(\lambda) =k=\mbox{diag}(1,\ 1,\ -1) \ \label{dk} \ee and expand
${\mathbb T}(\lambda)$, given by \be {\mathbb T}_0(\lambda) =
L_{0N}(\lambda)
 \ldots L_{01}(\lambda)\ k\ L^{-1}_{01}(-\lambda) \ldots L^{-1}_{0N}(-\lambda)\ ,
\ee
as (see also \cite{doikouy})
\be {\mathbb T}(\lambda) = k +
\sum_{k=1}^{2N}{{\mathbb T}^{(k-1)} \over \lambda^{k}} \ ,
\label{exppa6}
\ee
where
\be {\mathbb T}^{(0)} & = & i \sum_{i=1}^N
\Big ( {\mathbb P}_{0i}k + k {\mathbb P}_{0i} \Big )\ ,
\nonumber\\
{\mathbb T}^{(1)} & = & - \sum_{i>j} {\mathbb P}_{0i}\ {\mathbb
P}_{0j}\ k - k\ \sum_{i<j} {\mathbb P}_{0i}\ {\mathbb P}_{0j} -
\sum_{i,j =1}^N {\mathbb P}_{0i}\ k\ {\mathbb P}_{0j} - k
\sum_{i=1}^N {\mathbb P}_{0i}^2 \ , \label{first}
\ee
for the first two terms.
Before we proceed let us first recall what happens when $k ={\mathbb
I}$. In this case the open transfer matrix enjoys the full $gl_3$
symmetry (see e.g. \cite{done, doikouy, doikoucrampe}) and from the
trace of ${\mathbb T}^{(1)}$ we obtain the quadratic Casimir. For
instance for one-``particle'' ($N=1$) we obtain:
\be
C =
tr\{{\mathbb P}^{2}\} &=& (e^{(1)})^2 + (e^{(2)})^2 + (e^{(3)})^2 +
J^{-(1)} J^{+(1)} + J^{-(2)} J^{+(2)} + \Lambda^{-} \Lambda^{+} \non\\
&+& J^{+(1)} J^{-(1)} + J^{+(2)} J^{-(2)} + \Lambda^{+} \Lambda^{-}\
. \label{cas1} \ee Let us now come back to the situation where $k$
is given by (\ref{dk}). Then \be {\mathbb T}^{(0)} = 2i
\sum_{j=1}^N\left(
\begin{array}{ccc}
e_j^{(1)} &J_j^{-(1)} & 0\\
J_j^{+(1)} &e_j^{(2)} & 0 \\
     0 & 0 &-e_j^{(3)} \\ \end{array} \right )\ . \label{tt0}
\ee
Clearly, what remains consists of the $sl_2 \otimes u(1)$ algebra. Specifically,
$\big(\displaystyle \sum_j J^{\pm(1)},\ \sum_j s_j^{(1)}\big)$
satisfies the $sl_2$ commutation relations, whereas $\displaystyle \sum_i e_i^{(3)}$
commutes with everything.
The first two conserved quantities are given by taking the trace over the auxiliary space
in ${\mathbb T}^{(0)},\ {\mathbb T}^{(1)}$. We obtain
\be
t^{(0)} \propto \sum_{j=1}^N \Big ( c_j- 2 e_j^{(3)} \Big )\ ,
\quad \mbox{where} \qq c_j= e_j^{(1)}+ e_j^{(2)} + e_j^{(3)}
\ee
and
\be
&& t^{(1)} \propto \sum_{j=1}^N
\Big ( (e_j^{(1)})^2 + (e_j^{(2)})^2 - (e_j^{(3)})^2 + J_i^{-(1)}J_i^{+(1)} + J_i^{+(1)} J_i^{-(1)}
-c_j + 3 e^{(3)}_j \Big )
\nonumber\\
&&
\phantom{xxx}
+ 2\sum_{i< j} \Big ( e_i^{(1)}e_j ^{(1)}+ e_i^{(2)}e_j ^{(2)} - e_i^{(3)}e_j ^{(3)} + J_i^{-(1)}J_j^{+(1)} + J_i^{+(1)}
J_j^{-(1)}\Big )\ .
\label{kjnb2}
\ee
The one ``particle'' Hamiltonian $t^{(1)}$ is then
\be
t^{(1)} \propto (e^{(1)})^2 + (e^{(2)})^2 - (e^{(3)})^2 + J^{-(1)}J^{+(1)} + J^{+(1)} J^{-(1)}\ ,
\ee
where we have omitted the linear in the generators terms corresponding to
the first line of \eqn{kjnb2} since they commute with everything.
For notational convenience, we set
\be
s^{(1)}\equiv 2 J\ , \qq e^{(3)} \equiv 2 \tilde J\ , \qq J^{-(1)} \equiv -J^-\ , \qq J^{+(1)}\equiv J^+\ .
\ee
Then one finds
\be
t^{(1)} \propto J^2 -{1\over 2}\{J^{+},\ J^{-}\} -\tilde J^2 -c\tilde J\ .
\ee
From the first integral of
motion it is clear that $e^{(3)} = 2 \tilde J$ is also a conserved quantity, so the second charge may be written as
\be
I^{(0)} = \tilde c - 2 \tilde J\ , \qq I^{(1)}= J^2 -{1\over 2}\{J^{+},\ J^{-}\} -\tilde J^2\ \qq \tilde c = e^{(1)} + e^{(2)}
\label{1part}
\ee
$\tilde c$ is obviously a $sl_2$ central element.

Similarly, the $N$-tensor representations are given, respectively, by
\be
{\mathbb I}^{(0)}&=& \sum_{j=1}^N (\tilde c_j-2\tilde J_j)\ ,
\nonumber\\
 {\mathbb I}^{(1)} &=& \sum_{j=1}^N \Big (J_j^2 -{1\over 2}\{J_j^{+},\ J_j^{-}\} -\tilde J_j^2 \Big )
+ 4 \sum_{i< j}\Big( J_i J_j -{1\over 2}(J_i^{+}J_j^{-}+J_i^{-}J_j^{+}) -\tilde J_i \tilde J_j \Big )\ .
\label{npart}
\ee

\no
We shall exploit the breaking of the symmetry due to the presence of non-trivial integrable boundaries to obtain the centrally
extended $E_{2}^c$ algebra.
Consider the following contraction\footnote{This contraction is known in the mathematics
literature as a Saletan contraction and
is distinct from the Inon\"u--Winger contraction.
It is the analog of so called Penrose limit in gravity,
that constructs a plane
wave starting from any gravitational
background by magnifying the region around a null geodesic and
it was first used in the string literature in WZW models \cite{sfetsos}.
More recently, the Penrose limit has been taken in various supersymmetric brane solutions
of string and M-theory \cite{BFHPPenrose} and has been instrumental
in understanding issues within the AdS/CFT correspondence involving sectors of large quantum
numbers \cite{BMN}.
}
\be
J^{\pm} = {1 \over \sqrt{2\epsilon}} P^{\pm}\ , \qq J ={1 \over 2}\left(T+{F\over \epsilon}\right)\ ,
\qq \tilde J = - {F \over 2 \epsilon}\ ,\qq \e\to 0\ .
\label{map}
\ee
Then one obtains the following commutation relations that define the $E_{2}^c$ algebra
\be
[P^+,\ P^- ] = -2F\ , \qq [T,\ P^{\pm}] = \pm P^{\pm}\ ,
\label{ec2}
\ee
where $F$ is an {\it exact} central element of the algebra.
It is obvious that the conserved quantities, for the one-site case,
after contracting and keeping the leading order contribution are
\be
I^{(0)} = F\ , \qq I^{(1)} = T F-{1
\over 2} \{P^+,\ P^- \}\ .
\label{part1}
\ee
The $N$-site representation follows immediately from (\ref{npart})
\be
{\mathbb I}^{(1)} \sim \sum_{i=1}^N \Big ( F_i T_i-{ 1 \over 2}\{P_i^+,\ P^-_i\}\Big )
+ 2 \sum_{i<j} \Big (F_i T_j + T_i F_j - \{ P^+_i,\ P^-_j \} \Big )\ .
\label{e2c}
\ee

Let us now focus on the $N=1$ and see more precisely how one obtains higher Casimir operators of
$E_2^c$ from the expansion of the transfer matrix $t(\lambda) = \sum_{k=1}^{2N} {t^{(k-1)} \over \lambda^k}$.
Recall the $N=1$ representation of the reflection algebra
\be
&& {\mathbb T}(\lambda) = L(\lambda)\ k\ \hat L(\lambda) = (1 +{i\over \lambda}{\mathbb P})\ k\
(1 +{i\over \lambda}{\mathbb P} -{1\over \lambda^2}{\mathbb P}^2
-{i \over \lambda^3}{\mathbb P}^3 + {1\over \lambda^4}{\mathbb P}^4 \ldots) \non\\
&& = k + {i\over \lambda}({\mathbb P} k + k {\mathbb P}) -{1 \over \lambda^2}({\mathbb P} k {\mathbb P} +
k{\mathbb P}^2) - {1\over \lambda^2} ({\mathbb P}k {\mathbb P}^2 + k {\mathbb P}^3 ) \ldots \label{exp11}
\ee
where $k$ is given in (\ref{dk}) then
\be
t^{(k-1)} \propto \sum_{a, b}({\mathbb P}_{ab}\ k_{bb}\ {\mathbb P}^{k-1}_{ba} + k_{aa}\
{\mathbb P}_{aa}^k  ) \label{tk}
\ee
Before contraction $t^{(k)}$ are the higher Casimir operators of $sl_2 \otimes u_1$, after
contracting one has to consistently keep only the highest order
contribution in the ${1 \over \epsilon}$ expansion of each $t^{(k)}$. Then each one of $t^{(k)}$
commutes by construction with $E_2^c$. We have already explicitly computed the quadratic
 one (\ref{part1}), and
the derivation of higher Casimir $t^{(k)}$ is then simply a matter of involved algebraic computations,
since the generic form is known (\ref{tk}). It is clear that since each one of $t^{(k)}$
commutes with $E_2^c$ the transfer matrix also commutes. This logic may be generalized
for performing contraction to any higher rank algebra
or $q$ deformed algebra (see next section), the same applies for generic $N$. In fact, this argument
holds independently of the context one realizes
the contraction (see e.g. \cite{sfetsos}). More precisely, having in general a
set of Casimir operators of say the $gl_n$ algebra
after contraction one consistently should keep the highest order
contribution in order to obtain the
contracted Casimir quantities. Depending on the rank of the considered algebra
the expansion of $t(\lambda)$ should truncate at some point -note that expressions (\ref{exp11}), (\ref{tk})
are generic and hold for any $gl_n$-
or in other words the higher
Casimir quantities should be trivial combinations of the lower ones.
This is a quite intricate technical point,
however is beyond the scope of the
present article.

\no We may obtain a construction similar to the one above by
re-parametrizing the $L$ operator --solution of the fundamental
equation (\ref{fundam}), with $R$ being the ${\cal Y}(sl_2)$. Define
below the $L,\ \hat L$ operators
\be L(\lambda) = 1+{i\over
\lambda}\ {\mathbb P}\ , \qq \hat L(\lambda) = 1 +{i \over \lambda}\
\hat {\mathbb P}\ , \ee where \be {\mathbb P} =\left(
\begin{array}{cc}
\tilde J+ J &-J^- \\
J^+ & \tilde J-J \\ \end{array} \right) \ ,\qq \hat {\mathbb P} =\left(
\begin{array}{cc}
-\tilde J+ J +1 &-J^- \\
J^+ & -\tilde J-J +1 \\ \end{array} \right) \,.
\label{psinp}
\ee
To obtain a Casimir like quantity it is more natural to consider the open spin chain system
\be
{\mathbb T}_0(\lambda) = L_{0N}(\lambda) \ldots L_{01}(\lambda)\
 \hat L_{01}(\lambda) \ldots \hat L_{0N}(\lambda)\ .
\ee The charges in involution may be again obtained via appropriate
expansion. We omit here the relevant details for brevity. Finally,
after expanding and contracting we obtain from the $1/\lambda^2$
term
\be
 {\mathbb I}^{(1)} &=& \sum_{j=1}^N \Big (J_j^2 -{1\over 2}\{J_j^{+},\ J_j^{-}\} -\tilde J_j^2 \Big )
+ 4 \sum_{i< j}\Big( J_i J_j -{1\over 2}(J_i^{+}J_j^{-}+J_i^{-}J_j^{+}) \Big )\ .
\label{npart3}
\ee
After the contraction we obtain for the first non-trivial one-particle charge
\be
I^{(1)} = T F -{1\over 2} \{P^+,\ P^- \}\ ,
\label{part2}
\ee
whereas the $N$-particle charge becomes
\be
{\mathbb I}^{(1)} \sim \sum_{i< j} F_{i} F_j\ .
\label{npart2}
\ee

\no
Notice that the one ``particle'' conserved quantities (\ref{part1}) and
(\ref{part2}) coincide, whereas the $N$ ``particle'' charges (\ref{e2c}) and
(\ref{npart2}) are different.
In fact, the underlying symmetries in the two descriptions are different.
In the second case the remaining symmetry after implementing the boundary
is the $sl_2$ (the $u(1)$ symmetry is hidden)
whereas in the first description
it is the $sl_2 \otimes u(1)$.
The description of (\ref{npart}) gives the expected co-product associated
to the Casimir of the $sl_2 \otimes u(1)$,
so that in this sense it is a more natural description.

\section{The $U_q(E_{2}^c)$ algebra from $U_q(sl_2) \otimes u(1)$}

We shall now turn and study in our context the $q$-deformed situation.
More precisely, we shall focus on the construction of the centrally
extended $U_q(E_{2}^c)$ algebra from $U_q(sl_2) \otimes u(1)$. It was shown in \cite{done1}
that applying special boundary conditions in an open $U_q(gl_n)$ spin chain
(in the fundamental representation) breaks the symmetry to
$U_q(gl_l) \otimes U_q(gl_{n-l})$, where $l$ is an integer associated to the choice of boundary.
This statement was generalized for generic algebraic objects, independently of the choice of representation
(about boundary quantum algebras and for generic boundary conditions see \cite{myhecke, doikoutwin}).
The case with no central extension, that is $U_q(E_2)$,
has been obtained in earlier works \cite{qe2} from the $U_q(sl_2)$ algebra via a
Inon\"u--Wigner type contraction.

\no
Before we proceed let us first review some relevant facts regarding
the $U_q(gl_n)$ algebra \cite{jimbo}-\cite{cha}. Let
\be
a_{ij} = 2 \delta_{ij} - (\delta_{i\ j+1}+ \delta_{i\ j-1})\ , \qq i,j = 1,2, \ldots , n-1 \ ,
\ee
be the Cartan matrix of the Lie algebra $sl_n$.
The quantum enveloping algebra $U_{q}(sl_n)$ with the
Chevalley--Serre generators \cite{jimbo, drinf}
\be
e_{i}\ , \quad f_{i}\ ,\quad q^{\pm {s_{i}\over 2}}\ , \qq
i=1,2, \ldots, n-1\ ,
\ee
obey the defining relations
\be
&& \Big [q^{\pm {s_{i}\over 2}},\ q^{\pm {s_{j}\over 2}} \Big]=0\, \qquad q^{{s_{i}\over 2}}\ e_{j}=q^{{1\over
2}a_{ij}}e_{j}\ q^{{s_{i}\over 2}}\, \qquad q^{{s_{i}\over 2}}\ f_{j}
= q^{-{1\over 2}a_{ij}}f_{j}\ q^{{s_{i}\over 2}}\ ,
\non\\
&& \Big [e_{i},\ f_{j}\Big ] = \delta_{ij}{q^{s_{i}}-q^{-s_{i}} \over q-q^{-1}}\ ,
\qq i,j = 1,2, \ldots,n-1\ .
\label{1}
\ee
They also satisfy the $q$-deformed Serre-relations, omitted here for brevity (see e.g. \cite{jimbo}).
The above generators form the $U_{q}( sl_n)$ algebra and also,
$q^{\pm s_{i}}=q^{\pm (\epsilon_{i} -\epsilon_{i+1})}$. The $U_{q}(gl_n)$ algebra is derived
by adding to $U_{q}( sl_n)$ the elements $q^{\pm \epsilon_{i}}$ $i=1, \ldots, n$
so that $q^{\sum_{i=1}^{n}\epsilon_{i}}$ belongs to the center
(for more details see \cite{jimbo}).\footnote{We have changed the notation we have used
for the generators of the undeformed $gl_n$ case in order to conform with the literature we refer to.
The correspondence with them is
$e_i\to J^{+(i)}$, $f_i\to J^{-(i)}$, $\epsilon_i \to e^{(i)}$ and $s_i\to s^{(i)}$. In
the limit $q\to 1$ we recover from \eqn{1} the commutators \eqn{sllk}.
}

\no
The algebra $ U_{q}(gl_{n})$ is equipped with a co-product  $\Delta: U_{q}(gl_{n})
\to U_{q}(gl_{n})\otimes U_{q}(gl_{n})$, acting as
\be
\Delta(y) = q^{- {s_{i} \over 2}} \otimes y + y \otimes q^{{s_{i} \over 2}}\ , \qq
y \in \{e_{i},\ f_{i} \}\ ,\qq
 \Delta(q^{\pm{\epsilon_{i} \over 2}}) = q^{\pm{\epsilon_{i} \over 2}} \otimes
q^{\pm{\epsilon_{i} \over 2}}\ .
\label{cop}
\ee

\no We shall focus here in the (trigonometric) ${\mathbb
U}_q(\widehat{gl_n})$ $R$-matrix which is given by \cite{jimbo}
\be
&& R(\lambda) = a(\lambda) \sum_{i=1}^{n} \hat e_{ii} \otimes \hat
e_{ii}+ b(\lambda) \sum_{i\neq j=1}^{n} \hat e_{ii} \otimes \hat
e_{jj} + c \sum_{i\neq j =1}^{n} e^{ -sgn(i-j)\lambda} \hat
e_{ij}\otimes \hat e_{ji}\ , \ee where $R \in \mbox{End}(({\mathbb
C}^{n})^{\otimes 2})$ and $\hat e_{ij}$ are $n\times n$ matrices
with elements $(\hat e_{ij})_{kl} = \delta_{ik}\delta_{jl}$. We also
define for notational convenience \be a(\lambda) = \sinh \mu(\lambda
+i )\ , \qq b(\lambda) = \sinh \mu \lambda\ , \qq c = \sinh i\mu\
,\qq q= e^{i\mu}\ .
\ee
The associated $L$ operator in this case may
be written in the following form \cite{difre}
\be
L(\lambda) =
e^{\mu \lambda} L^+ - e^{-\mu \lambda} L^- \ , \label{ll}
\ee
where
\be
 L^+
= \sum_{i \leqslant j } \hat e_{ij} \otimes t_{ij}\ ,\qq L^- =
\sum_{i \geqslant j } \hat e_{ij} \otimes t_{ij}^-\ . \label{ll1}
\ee
For definitions of the elements $t_{ij}$ and $ t^-_{ij}$ we
refer to the Appendix.

\no As in the previous sections we are mostly interested in
representations of the reflection algebra (\ref{re2}), given in
(\ref{tt1}) and (\ref{mono}), where now $L$ is given by (\ref{ll}).
We also choose the homogeneity parameter $\Th =0$. To define
$L^{-1}(-\lambda)$ is quite intricate and in general is expressed in
powers of $e^{-\mu \lambda}$. For our purposes here we are interested in
the highest order of the expansion (as $\lambda \to \infty$)
\cite{myhecke}
\be
 L^{-1}(-\lambda) \sim \hat L^+ +{\cal
O}(e^{-2\mu \lambda})\ , ~~~\mbox{where} ~~~~~ \hat L^+ =
\sum_{i\geqslant j} \hat e_{ij} \otimes \hat t_{ij}\ .
\ee
For the
definition of $\hat t_{ij}$ we refer again to the Appendix as well.
One can readily check that $L^-\hat L^+=\mathbb{I}$ (with no
proportionality factor). The open transfer matrix is defined in
(\ref{transfer}), and for simplicity we shall set henceforth $K^{+}
\propto M$, where $M$ is a matrix that in the $U_q(gl_n)$ series -
in the homogeneous gradation (for details see e.g. \cite{done1,
myhecke})- has the form \be M = \sum_j q^{n-2j+1}\hat e_{jj}\ , \qq
\mbox{and} \qq [M_1 M_2,\ R_{12}(\lambda) ]=0\ . \label{jh5.10} \ee
Consider a $c$-number solution of the reflection equation of the
type (\ref{11}) then its asymptotic behavior as $\lambda \to \infty$
is $
K(\lambda \to \infty) \sim K^{(0)} + {\cal O}(e^{-2\mu \lambda})$.
The representation of the reflection algebra for one ``particle'' as
$\lambda \to \infty$, becomes ($T \to L$ and ${\mathbb T} \to
{\mathbb K}$)
\be
{\mathbb K}(\lambda \to \infty) = L^+\ K^{(0)}\
\hat L^+\ .
\ee
 We will only consider diagonal $c$-number solutions
of the diagonal boundary conditions that break $U_q(gl_{n})$ to
$U_q(gl_l) \otimes U_q(gl_{n-l})$, as was first shown in
\cite{done1}. If both $K^{+}, K \propto {\mathbb I}$ then the open
transfer matrix enjoys the full $U_q(gl_n)$ symmetry (see e.g.
\cite{done1, myhecke, doikoutwin} and references therein). In this
case, of trivial boundary conditions that preserve the full symmetry
one obtains the Casimir of the associated $U_q(gl_n)$ algebra.
Indeed as $\lambda \to \infty$ we obtain\footnote{Notice that as
$\lambda \to -\infty$ one obtains another Casimir operator for the
deformed algebra
\be
t^- = tr (M\ L^-\ \hat L^-) = \sum_{i\geqslant
j} q^{n -2i +1 } t^-_{ij} \hat t^-_{ji} \label{t-}\ .
\ee In the
$U_q(gl_3)$ case in particular it reduces to: \be t^- &=& q^2
q^{-2\epsilon_1} +q^{-2 \epsilon_2} + q^{-2}q^{-2\epsilon_3} +
(q-q^{-1})^2\left (q^{-\epsilon_1 - \epsilon_2 +1} e_1f_1 +
q^{-\epsilon_2 - \epsilon_3 -1}e_2 f_2\right ) \non\\ && +\
(q-q^{-1})^2q^{-\epsilon_1 -\epsilon_3}(q^{-1}e_1e_2- e_2e_1)(f_2f_1
-q f_1 f_2). \label{cas3}
\ee}
\be
t^+ =tr (M\ L^+\ \hat L^+) =
\sum_{i\leqslant j} q^{n -2i +1 } t_{ij} \hat t_{ji}\ .
\ee
This is
perhaps the most natural and simplest way to obtain Casimir
operators of $q$ deformed algebras. For instance, in the case of
$U_q(gl_3)$, we obtain \be L^+ = \left(
\begin{array}{ccc}
q^{\epsilon_1} &t_{12} &t_{13}\\
 0 & q^{\epsilon_1} &t_{23} \\
0 & 0 &q^{\epsilon_1} \\ \end{array} \right )\ , \qq
\hat L^+ = \left(
\begin{array}{ccc}
q^{\epsilon_1} &0 &0 \\
\hat t_{21} &q^{\epsilon_1} &0 \\
\hat t_{31} &\hat t_{23} & q^{\epsilon_1} \\ \end{array} \right )\ .
\label{asy3}
\ee
Then, the explicit expression of the Casimir operator is given by
\be
t^+ &=& q^2 q^{2\epsilon_1} + q^{2\epsilon_2}
+ q^{-2} q^{2\epsilon_3} + (q-q^{-1})^2 \left( q^{\epsilon_1 + \epsilon_2+1} f_1 e_1 +
q^{\epsilon_2 + \epsilon_3-1}f_2 e_2\right)
\non\\
 && + \ (q-q^{-1})^2 q^{\epsilon_1 + \epsilon_3} (q f_2 f_1 - f_1 f_2)(e_1 e_2
- q^{-1} e_2 e_1)\ . \label{cas2}
\ee
The above expression is quite
compact and depends in a straightforward manner on the algebra
generators (see also \cite{lizh}). Notice that by construction the
quantity $t^{+}$ belongs to the center of $U_q(gl_3)$. Recall that
in this case the transfer matrix enjoys the full $U_q(gl_3)$
symmetry \cite{done1, myhecke}, thus all the charges in involution
generated from the transfer matrix belong to the center of
$U_q(gl_3)$. Note that, using (\ref{cas2}) and (\ref{cas3}), the sum
of $t^+$ and $t^-$ reduces to the $gl_3$ quadratic Casimir
(\ref{cas1}) in the isotropic limit $q \to 1$. In general the
spectrum of these Casimir type quantities associated to any
$U_q(gl_n)$ invariant open spin system may be readily computed via
Bethe ansatz techniques for any representation. The spectrum and
Bethe ansatz equations are known for these open spin chains
\cite{annecy3}. Hence, expanding it in powers of $e^{-\lambda}$ will
provide the eigenvalues of the associated Casimir operators.

\no
We are basically interested in obtaining $U_q(E_{2}^c)$ as a
contraction of the $U_{q}(sl_2) \otimes u(1)$, which again is a boundary symmetry.
We shall focus henceforth on $U_q(gl_3)$ and on diagonal $K$-matrices of the form \cite{deve}
\be
K(\lambda) = \ \mbox{diag} \Big (e^{\mu \lambda},\ e^{\mu \lambda},\
-e^{-\mu \lambda}\Big )\ . \label{11} \ee We take into account the
asymptotics $L^+$, $\hat L^+,$ in the $U_q(gl_3)$ (\ref{asy3}) and
$K^{(0)} =\mbox{diag} (1, \ 1,\ 0)$. We may now explicitly write \be
{\mathbb K} (\lambda \to \infty) \sim {\mathbb K}^+ = \left(
\begin{array}{ccc}
q^{2\epsilon_1} + t_{12} \hat t_{21} &t_{12}q^{\epsilon_2} & 0 \\
q^{\epsilon_2}\hat t_{21} &q^{2 \epsilon_2} &0 \\
  0 & 0 &0 \\ \end{array} \right ) \ ,
\label{kk}
\ee
where using the Appendix we have
\be
t_{ii} = q^{\epsilon_i}\ , \qq t_{12} =(q- q^{-1}) q^{-1/2} q^{{\epsilon_1 +\epsilon_2 \over 2}} f_1\ ,
\qq
\hat t_{21} \equiv (q- q^{-1}) q^{-1/2} q^{{\epsilon_1 +\epsilon_2 \over 2}} e_1\ .
\ee
It is also convenient to implement the following identifications
\be
e_1 \equiv J^+\ , \qq f_1 \equiv -J^-\ , \qq \epsilon_1-\epsilon_2 \equiv 2 J\ ,
\qq \epsilon_1 + \epsilon_2 = - 2 \tilde J\ .
\ee
clearly $\tilde J$ is central element of $U_q(sl_2)$.
The asymptotic expression of the transfer matrix as $\lambda \to \infty$
provides the first conserved quantity (higher terms in the expansion give rise to higher charges)
\be
t^+ = tr \{M {\mathbb K}^+ \}\ ,
\ee
where from \eqn{jh5.10} the matrix $M = \diag(q^2,1,q^{-2})$.
We conclude that
\be
t^+ \propto q^{-2\tilde J} \Big ( q^{2J+1} + q^{-2J-1} - (q-q^{-1})^2 J^-J^+ \Big )\ ,
\label{cas}
\ee
which is the Casimir operator of $U_q(sl_2)$ and $q^{-\tilde J}$ is apparently central element of $U_q(sl_2)$.
This is somehow expected given that the associated transfer matrix enjoys the
exact $U_q(sl_2) \times u(1)$ symmetry (see,
e.g. \cite{done1, myhecke, doikoutwin} for details on the proof).
The $u(1)$ charge is obtained from $\lambda \to -\infty$ asymptotic behavior
of the transfer matrix
and it is (we omit the technical details for brevity)
\be
t^- = \epsilon_3.
\ee
In the $N$-tensor representation one obtains as expected the following non-local quantities
\be
&& t^{+(N)} \propto \Delta^{(N)}(q^{-2\tilde J})
\Big (q \Delta^{(N)}(q^{2J}) + q^{-1} \Delta^{(N)}(q^{-2J}) - (q-q^{-1})^2
\Delta^{(N)}(J^-)\Delta^{(N)}(J^+) \Big )\ ,
\non\\
&& t^{-(N)} =\Delta^{(N)}(\epsilon_3)
\ee
where the indicated $N$ co-products are derived via (\ref{cop}) by iteration. Higher order Casimir operators
of the $q$-deformed algebra are obtained by considering the expansion of the open transfer matric
in powers of $e^{\pm 2 \mu \lambda}$;
the same arguments -for the contracted version as well- hold
as in the rational case described in the previous sections.

\no
Consider the Saletan-type contraction \eqn{map} after also setting
the deformation parameter to $q = e^{\epsilon \eta}$. This will lead to the deformed centrally
extended $U_q(E_{2}^c)$ algebra. We obtain
\be
[T,\ P^{\pm}] = \pm 2 P^{\pm}\ , \qq [P^+,\ P^-] = -2 F_\eta\ , \quad F_\eta = {\sinh (\eta F) \over \eta}\ ,
\label{e2ceta}
\ee
where $F_\eta$ is an {\it exact} central element of the algebra. The associated co-products
emanating from (\ref{cop}) are given by
\be
&& \Delta(P^{\pm}) = e^{-{\eta F\over 2}} \otimes P^{\pm} + P^{\pm} \otimes q^{{\eta F \over 2}},
\non\\
&&\Delta(e^{\pm \eta F}) =e^{\pm \eta F} \otimes e^{\pm \eta F},
\label{cop2}\\
&& \Delta(T) = {\mathbb I} \otimes T + T \otimes {\mathbb I}\ .
\non
\ee
Note that, although the algebra \eqn{e2ceta} is an $E_{2}^c$ one for the central extension $F_\eta$,
what appears in the co-products is $F$ itself.

\no
The associated Casimir follows from (\ref{cas}) after we perform the contraction.
We find that
\be
C= e^{\eta F} \Big (2 \cosh(\eta F) + 2 \eta^2 \epsilon
( T F_\eta - {1\over 2} \{P^+,\ P^-\}) \Big ) \ .
\ee
It is straightforward to check that the latter quantity commutes with all the
elements of the algebra $U_q(E_{2}^c)$.
The $N$-co-product Casimir follows from (\ref{cop2}) by iteration. Given that $F$
is a central element of the constructed algebra we
conclude that
\be
I_{\eta} = T F_\eta - {1\over 2} \{P^+,\ P^-\}\ ,
\ee
is also a conserved quantity.
It is clear that in the isotropic limit $\eta \to 0 $ the algebra reduces to \eqn{ec2}
and the associated Casimir operator to \eqn{part2}.

\section{Discussion}

The main theme of the present paper is that by exploiting the symmetry breaking mechanism due to the presence
of integrable boundary conditions, one can naturally
construct contracted and centrally extended algebras.

\no
In order to simplify
the analysis and clearly demonstrate the main ideas we restricted our discussion to examples involving
originally $sl_2$ or $gl_3$ symmetry.
A natural extension of the present work is to consider generic symmetry breaking of the type
${\mathrm G} \otimes {\mathrm H}$ where ${\mathrm G}$ and
${\mathrm H}$ are generic algebras (${\mathrm H} \subset {\mathrm G}$),
and then follow a contraction procedure similar to the ones for ordinary Lie and current algebras
(see e.g. \cite{sfetsos}).
Also, note that the generic study of boundary symmetry breaking for higher rank algebras is presented in
\cite{done, done1, myhecke}, and the super-symmetric case is a work in progress at the moment.

\no
We were able to obtain the twisted Yangian of $E_2$ as well as the centrally extended $E_2^c$ and
$U_q(E_2^c)$ algebras via suitable contractions of $gl_3$ and $U_q(gl_3)$ algebras.
Naturally, one may wonder if one could have
started the whole analysis by directly considering the $E_2$, $E_2^c$ and $U_q(E_2^c)$ algebras.
However, such an approach would require
knowledge of the universal $R$-matrices associated to $E_2$, $E_2^c$ and $U_q(E_2^c)$.
In general, such a derivation is an intricate issue,
so the approach we followed here is admittedly the most straightforward and simplest one.
Nonetheless it is clear according to the original works in \cite{drinf, jimbo, tak} that
linear exchange relations between the universal R-matrix and co-products of the charges
of the Yangian or  the affine $q$-algebra (see expressions of non-local charges in
(\ref{first0}), (\ref{cc}), (\ref{c2}), (\ref{first}), (\ref{tt0})) provide
the exact form of the $R$-matrix. So our results
may in addition be utilized for exactly deriving the universal $R$-matrix associated
to the the Yangian of $E_2$, $E_2^c$ or the $q$-deformed $E_2^c$.

\no
A more ambitious task is to explore and possibly apply our methods
in the context of the AdS/CFT correspondence, where certain quantum group structures and
supersymmetric centrally extended algebras arise
(see, for instance, \cite{beisert,GomHer,Plefka}).
The pertinent question is wether one can use the
generic context described here in order to uncover the full underlying algebraic structure,
that is to extract by
the methodology proposed the associated centrally extended algebra.
That would be
very important in understanding fundamental aspects of the AdS/CFT correspondence finitely
beyond the perturbative level.

\no
We hope to address the aforementioned significant issues in future works.

\appendix

\section{Appendix}

In this appendix we shall introduce some useful quantities (elements of $U_q(gl_n)$).
Define ${\cal E}_{i\ i+1} = \hat {\cal E}_{i\ i+1}= e_{i}$ and ${\cal E}_{i+1\ i}
= \hat {\cal E}_{i+1\ i} = f_i$,
$i= 1, 2,\ldots, n-1 $ and for $|i-j|>1$, whereas
\be
&& {\cal E}_{ij} = {1\over |i-j|-1} \sum_{k={\rm min}(i,\ j)+1}^{max(i,\ j)-1}( {\cal E}_{ik}\
 {\cal E}_{kj} -q^{\mp 1} {\cal E}_{kj}\ {\cal E}_{ik})\ ,\qq j \lessgtr k \lessgtr i \ ,
\non\\
&&\hat {\cal E}_{ij} = {1\over |i-j|-1}
 \sum_{k={\rm min}(i,\ j)+1}^{{\rm max}(i,\ j)-1}(\hat {\cal E}_{ik}\ \hat {\cal E}_{kj}
-q^{\pm 1}\hat {\cal E}_{kj}\ \hat {\cal E}_{ik})\ , \qq j\lessgtr k \lessgtr i\ ,
\non\\
 && i,j =1, 2, \ldots , n\ .
\label{q1}
\ee
Also define
\be
&&t_{ij} = (q-q^{-1}) q^{-{1\over 2}} q^{{\epsilon_{i} \over 2}}q^{{\epsilon_{j} \over 2}}\
 {\cal E}_{ji}\ , \quad i<j\ , \quad t^{-}_{ij} =
-(q-q^{-1}) q^{{1\over 2}} q^{-{\epsilon_{i} \over 2}}q^{-{\epsilon_{j} \over 2}}\ {\cal E}_{ji}\ ,
 \quad i>j
\non\\
&&\hat t_{ij}= (q-q^{-1}) q^{-{1\over 2}} q^{{\epsilon_{i} \over 2}}q^{{\epsilon_{j} \over 2}}\
 \hat {\cal E}_{ji}\ , \quad i>j\ ,
\quad
 \hat t^{-}_{ij} = -(q-q^{-1}) q^{{1\over 2}} q^{-{\epsilon_{i} \over 2}}q^{-{\epsilon_{j} \over 2}}\
\hat{\cal E}_{ji}\ , \quad i<j\ ,
\non\\
&& t_{ii} =\hat t_{ii}=(t_{ii}^{-})^{-1} =(\hat t_{ii}^{-})^{-1}=q^{\epsilon_{i}}\ .
\label{q2}
\ee

\end{document}